\documentclass[11pt,a4paper]{article}
\pdfoutput=1
\usepackage{jcappub}

\usepackage[T1]{fontenc}

\usepackage{booktabs}
\usepackage{multirow}
\usepackage[dvipsnames]{xcolor}
\usepackage{array}
\usepackage{booktabs}
\usepackage{multirow}
\usepackage{xltabular}

\title{\boldmath Raising the Optical Depth to Reionization with Dark Matter Decay}

\author[a,b]{Yu-Ning Wang,}
\author[c]{Paulo Montero-Camacho,}
\author[c]{Ziwei Wang,}
\author[c]{Yin Li,}
\author[a,b]{and Yue-Lin Sming Tsai}

\affiliation[a]{Key Laboratory of Dark Matter and Space Astronomy,
Purple Mountain Observatory, Chinese Academy of Sciences,\\
Nanjing 210033, China}

\affiliation[b]{School of Astronomy and Space Science,
University of Science and Technology of China,\\
Hefei, Anhui 230026, China}

\affiliation[c]{Department of Strategic and Advanced Interdisciplinary Research,
Peng Cheng Laboratory,\\
Shenzhen, Guangdong 518000, China}

\emailAdd{ynwang@pmo.ac.cn}
\emailAdd{pmontero@pcl.ac.cn}
\emailAdd{ziweiwang13@gmail.com}
\emailAdd{eelregit@gmail.com}
\emailAdd{smingtsai@pmo.ac.cn}

\abstract{
Recent DESI DR2 baryon acoustic oscillation measurements have sharpened a discrepancy with cosmic microwave background (CMB) inferences that can be recast as a preference for a reionization optical depth of $\tau_{\rm reio}\simeq 0.09$, well above the $\simeq 0.06$ typically inferred from large-scale CMB polarization measurements.
Here we test whether electromagnetic energy injection from dark matter (DM) decaying through $\chi\rightarrow e^{+}e^{-}$ can provide this additional optical depth.
Adopting a Gompertzian model for astrophysical reionization, which provides an asymmetric, simulation-calibrated description of cosmic reionization, we derive $\tau_{\rm reio}$ self-consistently from the resulting evolution of the free electron fraction, including the contribution from DM decay.
We jointly constrain the cosmological and reionization parameters and the DM decay rate  using the latest ACT+Planck CMB temperature and polarization anisotropies, CMB lensing, BAO measurements, and quasar damping-wing observations.
For $m_\chi=1\,{\rm GeV}$, DM decay raises the marginalized optical depth from $\tau_{\rm reio}=0.064$ to $0.072$ with a flat prior on the decay rate, but does not significantly improve the fit. Reweighting the existing chains with log-uniform priors yields lower mean optical depth, indicating that the inferred increase depends on the assumed prior.
The additional optical depth arises primarily from a broad ionization tail extending through the cosmic dark ages, rather than from an earlier onset of astrophysical reionization.
Across $m_\chi=2\,{\rm MeV}$--$7\,{\rm GeV}$, the central $68\%$ credible intervals remain below the $\tau_{\rm reio}\simeq 0.08$ level previously found sufficient to bring both the neutrino-mass tension and the exclusion of $\Lambda$CDM below the $2\sigma$ level.
The resulting DM lifetime constraints are complementary to bounds from photon and cosmic-ray observations and gas heating, and are competitive with previous CMB limits.
Decaying DM therefore shifts the inferred optical depth in the direction required to ease the CMB--BAO  discrepancy, but the allowed increase remains insufficient to alleviate it substantially.
}


\begin{document}
\maketitle
\flushbottom

\section{Introduction}

The three-year Dark Energy Spectroscopic Instrument (DESI) Data Release 2 (DR2) baryon acoustic oscillation (BAO) measurements have sharpened the tension between BAO-inferred and cosmic microwave background (CMB)-inferred parameters within flat $\Lambda$CDM, increasing from 1.9$\sigma$ in DR1 to 2.3$\sigma$ in DR2~\cite{tr6y-kpc6}.
This discrepancy reflects DESI's preference for a lower matter density, $\Omega_m\simeq 0.2975$ compared with $\Omega_m \simeq 0.315$ from Planck~\cite{2020A&A...641A...6P}.

Within $\Lambda$CDM, this tension can be expressed in three closely equivalent ways~\cite{6vd2-rbfn}: as a preference for the sum of neutrino masses below the oscillation lower bound~\cite{Craig:2024tky, w9pk-xsk7, 613p-pph2}, as an excess CMB lensing amplitude~\cite{PhysRevD.111.083507}, or as an optical depth formulation of the same discrepancy~\cite{6r54-8lv4}. Each formulation reflects the same underlying discrepancy between the cosmological parameters inferred from CMB and BAO data.

Among these equivalent descriptions, the optical depth formulation is the most directly useful for constructing a physical model. 
The neutrino mass preference cannot be interpreted literally, since neutrino oscillation experiments impose a nonzero minimal mass~\cite{Esteban:2024eli}.
The lensing amplitude formulation is often captured by an artificial rescaling of the CMB lensing power and does not correspond to a physical parameter within $\Lambda$CDM.
By contrast, the optical depth $\tau_{\rm reio}$ is a physical observable directly determined by the abundance of free electrons, and thus the CMB--BAO discrepancy can potentially be addressed by a change in the evolution of the free electron fraction.

Large-scale CMB polarization measurements constrain $\tau_{\rm reio}$ and help calibrate the primordial amplitude $A_s$ by breaking the $A_s e^{-2\tau_{\rm reio}}$ degeneracy present in the small-scale CMB intensity acoustic peaks~\cite{PhysRevD.92.123535}.
A higher $\tau_{\rm reio}$ therefore implies an earlier or more extended reionization history and a higher inferred $A_s$, which increases the expected CMB lensing power and affects cosmological constraints on the sum of the neutrino masses $\sum m_\nu$~\cite{6vd2-rbfn,6r54-8lv4}.

Reframing the CMB--BAO discrepancy in terms of the evolution of the free electron fraction allows the apparent $\sum m_\nu$ tension to be reinterpreted as a preference for a higher optical depth.
The key issue is therefore whether a physically allowed free electron history can yield the required increase in $\tau_{\rm reio}$ while remaining consistent with observational constraints.
The value suggested by this optical-depth interpretation, $\tau_{\rm reio}\simeq 0.09$~\cite{6vd2-rbfn,6r54-8lv4}, is substantially above the standard Planck value, $\tau_{\rm reio}=0.054$, inferred from large-scale polarization under conventional reionization assumptions~\cite{2020A&A...641A...6P}. 
The same increase in $\tau_{\rm reio}$ also reduces the preference for evolving dark energy found in joint analyses of DESI BAO and CMB data under extended cosmologies, from $\simeq 3\sigma$ to $\simeq 1.5\sigma$~\cite{6r54-8lv4}.
Thus, raising $\tau_{\rm reio}$ is not merely a shift in a phenomenological parameter; it requires a physically motivated description of the reionization history and, potentially, an additional source of ionization. At the same time, this extra optical depth cannot simply be absorbed into an unconstrained high-redshift ionized component.
Flexible free electron histories can allow additional ionization at $z>15$, but improved large-scale polarization data do not provide significant evidence for a substantial early component~\cite{2018A&A...617A..96M,PhysRevD.104.063505}.
The CMB-inferred optical depth also remains close to $\tau_{\rm reio}\simeq 0.06$ in broad extensions of the standard cosmological parameter space. 
Analyses that simultaneously vary, among other parameters, the running of the scalar spectral index, dynamical dark energy, the sum of neutrino masses, and the scaling of the CMB lensing amplitude consistently find $\tau_{\rm reio}\simeq 0.058$~\cite{Roy_Choudhury_2024,Roy_Choudhury_2025}, suggesting that the low inferred value is not simply an artifact of restricting the analysis to the standard six-parameter $\Lambda$CDM model.

These results indicate that obtaining a larger optical depth is not straightforward within observationally allowed free electron histories; however, the flexible history analyses discussed above rely on generic functions to parametrize the evolution of the free electron fraction.
This motivates the use of a physically motivated model for astrophysical reionization.
The conventional tanh prescription~\cite{2008PhRvD..78b3002L}, widely used in CMB analyses, provides a convenient effective description but imposes a symmetric sigmoid form on the reionization transition.
Such a symmetric template is not well matched to astrophysical reionization: because reionization is driven by ultraviolet-emitting sources, its early phase should proceed gradually, followed by a more rapid completion once sources become abundant and ionized regions grow and overlap.
Indeed, semi-numerical reionization simulations (e.g., \texttt{21cmFAST}~\cite{10.1111/j.1365-2966.2010.17731.x}) show that the neutral hydrogen fraction exhibits an approximately universal asymmetric evolution, well described by the Gompertz mortality law~\cite{Montero-Camacho:2024dzs}.
The same universal shape also accurately describes the reionization history obtained from the THESAN-1 radiation-magnetohydrodynamic simulation~\cite{10.1093/mnras/stab3710}; see Figure 5 of~\cite{Montero-Camacho:2024dzs}.
Compared with the conventional tanh prescription, the Gompertzian reionization model based on this universal shape provides a more physically motivated description of the reionization process and aligns more naturally with current constraints on $x_{\rm HI}(z)$ from high-redshift quasars and galaxies~\cite{Montero-Camacho:2024dzs,Montero-Camacho:2026ohy}.

In Gompertzian reionization models, $\tau_{\rm reio}$ is not treated as a freely adjustable parameter but is instead derived from the reionization history specified by the model parameters.
When combined with CMB and Epoch of Reionization (EoR) observations, this physically constrained reionization history leads to substantially tighter constraints on $\tau_{\rm reio}$ and the primordial amplitude $A_s$ than the conventional tanh treatment~\cite{Montero-Camacho:2024dzs,Montero-Camacho:2026ohy}.
Thus, if alleviating the neutrino mass and lensing discrepancies requires $\tau_{\rm reio}\simeq 0.09$~\cite{6vd2-rbfn,6r54-8lv4}, it is nontrivial for observationally constrained astrophysical reionization alone to provide the required additional optical depth~\cite{Montero-Camacho:2026gaz,2026arXiv260924989U}.

If the additional optical depth cannot be supplied by astrophysical sources alone, then an additional physical process may contribute to the evolution of the free electron fraction before or during reionization.
To address the CMB--BAO discrepancy, such a contribution must satisfy two requirements: it must produce enough free electrons to increase $\tau_{\rm reio}$ appreciably, while remaining consistent with independent observational constraints provided by CMB and EoR observations.
Energy injection from dark matter (DM) provides a natural class of mechanisms, including both annihilation and decay.
However, previous studies have found that, once existing constraints are imposed, a sizeable contribution to the free electron fraction before astrophysical reionization is difficult to obtain from annihilation scenarios, while light DM decays into electromagnetic final states remain a more plausible possibility~\cite{PhysRevD.94.063507,PhysRevD.85.043522}.
We therefore focus on decaying DM.
Its electromagnetic decay products can deposit energy into the intergalactic medium, increasing the free electron fraction and heating the gas, thereby modifying the free electron history and the integrated optical depth~\cite{PhysRevD.94.063507} (related ionization effects from photons produced by axion decays were explored in earlier work~\cite{Berezhiani:1989fu}).
At the same time, CMB temperature and polarization anisotropies provide strong constraints on such energy injection~\cite{PhysRevD.70.043502, PhysRevD.95.023010}.
A related study investigated high-redshift ionization from Hawking evaporation of primordial black holes, finding that the allowed increase in optical depth was limited to $\Delta\tau\simeq0.008$ in their analysis~\cite{Sierra:2026vtz}.
In this work, we quantify the shift in the optical depth due to energy injection by DM decay allowed by current observational constraints.

The remainder of this paper is organized as follows.
Section~2 describes the construction of the free electron history, including the Gompertzian model for astrophysical reionization, energy deposition from DM decay, and the resulting optical depth and CMB observables.
Section~3 presents the statistical analysis and results. 
After introducing the parameters, priors, and data, we examine the representative $m_\chi=1\,\mathrm{GeV}$ case.
We then identify the origin of the additional optical depth, extend the analysis across the DM mass range, and assess the implications for the CMB--BAO discrepancy. 
Section~4 summarizes our conclusions.

\section{Free electron history with decaying dark matter}
\label{sec:ionization_model}
The optical depth depends on the free electron number density integrated along the line of sight and therefore records ionization throughout the post-recombination history. Astrophysical sources drive the reionization transition, while energy injection from DM decay provides an additional source of ionization during the preceding dark ages.
The effect of this injection depends on how the decay products propagate, cool, and deposit energy into ionization, excitation, and heating of the intergalactic medium (IGM). An extended, weakly ionized component can thus contribute to $\tau_{\rm reio}$ even without an earlier astrophysical reionization transition.

\subsection{Astrophysical reionization: the Gompertzian model}
\label{sec:gomp}

We adopt the Gompertzian model of Ref.~\cite{Montero-Camacho:2024dzs}, which describes the evolution of the global neutral hydrogen fraction found in both semi-numerical and fully coupled hydrodynamical reionization simulations.
The Gompertzian model is asymmetric and approximately universal in shape after a rescaling of the scale factor $a=(1+z)^{-1}$.
In the original construction, two shape parameters, $\alpha$ and $\beta$, are fitted jointly with the universal shape to the simulated histories; symbolic regression then relates these fitted parameters to the cosmological and astrophysical inputs.
Here we retain the universal shape and treat $\alpha$ and $\beta$ as free parameters.

The neutral hydrogen fraction is then written as
\begin{equation}
\begin{aligned}
x^{\rm Gomp}_{\rm HI}(\tilde a)
&= {\rm gomp}\!\left(R(\tilde a)\right)
\equiv \exp\!\left[-\exp\!\left(R(\tilde a)\right)\right],
\\
R(\tilde a)
&= \ln \tilde a\,
\frac{
1 + 0.1275 \ln \tilde a + 0.1022 \ln^2 \tilde a
}{
1 - 0.0060 \ln \tilde a + 0.0815 \ln^2 \tilde a
},
\\
\tilde a(a) &= \left[\frac{a}{\alpha}\right]^{\beta} .
% \ln \tilde a &= \beta\left(\ln a-\ln\alpha\right).
\end{aligned}
\label{eq:gomp}
\end{equation}
Here the parameter $\alpha$ sets the characteristic timing of reionization, while $\beta$ rescales the time variable and controls the width or steepness of the transition.

The corresponding hydrogen-sector free electron fraction is 
\begin{equation} 
    x_{e,{\rm H}}^{\rm Gomp}(z) = 1-x_{\rm HI}^{\rm Gomp}(z),
    \label{eq:xeH_gomp} 
\end{equation} 
which defines the astrophysical hydrogen baseline used throughout our analysis.

Joint analyses of CMB and astrophysical reionization observations using Gompertzian models restrict the allowed histories and tighten constraints on $\tau_{\rm reio}$, even after marginalizing over the relevant astrophysical uncertainties~\cite{Montero-Camacho:2024dzs,Montero-Camacho:2026ohy}.
They typically favor $\tau_{\rm reio}\sim0.05$--$0.06$, with little support for the substantially larger value $\tau_{\rm reio}\simeq0.09$ suggested by the optical-depth interpretation of the CMB--BAO discrepancy~\cite{6vd2-rbfn,6r54-8lv4}.
This motivates the additional ionization source described below.

\subsection{Dark matter decay and energy deposition}

For $s$-wave annihilation, the approximately velocity-independent cross section leads to substantial energy injection at high redshift, which is strongly constrained by CMB anisotropies~\cite{2020A&A...641A...6P,Slatyer:2015jla}.
For $p$-wave annihilation, the cross section is proportional to the squared DM velocity, suppressing injection in the smooth early Universe but allowing an enhancement as structure formation increases the density and velocity dispersion inside halos.
These models are less directly constrained by recombination-era CMB anisotropies, but face limits from late-time IGM heating and diffuse gamma-ray observations~\cite{PhysRevD.94.063507,Diamanti:2013bia}.

By contrast, energy injection from DM decay scales linearly with the DM density, $\rho_\chi$, making it less sensitive to nonlinear structure formation and allowing an extended ionization tail from the dark ages to the onset of astrophysical reionization.
Within the astrophysical and indirect-detection constraints considered in Ref.~\cite{PhysRevD.94.063507}, a contribution larger than ten percent to the pre-reionization free electron fraction is disallowed for both $s$-wave and $p$-wave annihilation, but remains possible for light DM decaying into electron--positron pairs.
We therefore adopt $\chi\to e^+e^-$ as our benchmark channel.\footnote{Ref.~\cite{PhysRevD.94.063507} motivates the choice of injection mechanism and channel. Its decay analysis does not provide a dedicated CMB anisotropy likelihood constraint; such constraints have been studied, for example, in Refs.~\cite{PhysRevD.95.023010,Diamanti:2013bia,Oldengott_2016}.}

We parameterize the decay model by the DM mass $m_\chi$ and the decay rate $\Gamma_\chi \equiv \tau_\chi^{-1}$, where $\tau_\chi$ is the DM lifetime.
For the benchmark channel $\chi\to e^+e^-$, the injected electromagnetic power per unit volume is
\begin{equation}
    \left(\frac{dE}{dVdt}\right)_{\rm inj} =
    \Gamma_\chi \rho_{\chi,0}(1+z)^3 ,
    \label{eq:dE_inj}
\end{equation}
where $\rho_{\chi,0}$ is the present-day DM energy density. 
For the parameter range considered in this work, $\tau_\chi$ is much longer than the age of the Universe (see Figure~\ref{fig:dm_decay_lifetime_posterior}), so the depletion of the DM abundance can be neglected.

The injected particles propagate and cool before depositing their energy into the gas.
Following Ref.~\cite{Slatyer:2015kla}, we encode this delayed deposition in redshift-dependent efficiencies.
We write the deposited power into a channel $\alpha$ as
\begin{equation}
    \left(\frac{dE}{dVdt}\right)_{\rm dep}^{\alpha}
    =
    f_\alpha(z,m_\chi)
    \left(\frac{dE}{dVdt}\right)_{\rm inj},
    \label{eq:dE_dep}
\end{equation}
where $\alpha$ denotes ionization, excitation, or heating.
Ionization directly increases the free electron abundance, while excitation, particularly Lyman-$\alpha$ excitation, can also contribute to the effective ionization rate by promoting neutral hydrogen atoms to states that are more easily ionized.
Heating modifies the gas temperature.

For each DM mass, the injected particle spectra are calculated with \texttt{Hazma}~\cite{Coogan_2020}, and their subsequent propagation and energy deposition are followed with \texttt{DarkHistory}~\cite{PhysRevD.101.023530}.
The Gompertzian hydrogen baseline in Eq.~\eqref{eq:xeH_gomp} is supplied to \texttt{DarkHistory}, since both the deposition and the subsequent ionization and thermal response depend on the ambient ionization state.
The background quantities are evaluated consistently with the cosmological parameters varied in the inference.
Because \texttt{DarkHistory} does not include a full treatment of massive neutrinos in the background evolution, we account for their contribution to $H(z)$ using the analytic fitting formula introduced in the WMAP Y7 analysis~\cite{2011ApJS..192...18K}.

From this calculation, we extract the DM-induced excess hydrogen-sector free electron fraction, $\Delta x_{e,{\rm H}}^{\rm DM}(z)$.
The modified hydrogen contribution is then written as
\begin{equation} 
    x_{e,{\rm H}}(z) = 
    x_{e,{\rm H}}^{\rm Gomp}(z) + \Delta x_{e,{\rm H}}^{\rm DM}(z). 
    \label{eq:xeH}
\end{equation}
The additive notation in Eq.~\eqref{eq:xeH} denotes an excess evaluated in the presence of the Gompertzian model contribution.
We extract only the DM-induced modification to the hydrogen sector from \texttt{DarkHistory}; the helium contribution is treated separately, as described below.

\subsection{Total free electron fraction, optical depth, and CMB observables}
\label{sec:cmb_calculation}

We construct the total free electron fraction by combining $x_{e,{\rm H}}(z)$ from Eq.~\eqref{eq:xeH} with the residual post-recombination electron fraction and helium contribution.
Following the convention of Ref.~\cite{Montero-Camacho:2024dzs}, the first ionization of helium is assumed to trace the modified hydrogen reionization history, while the second helium reionization at low redshift is modeled with a standard tanh transition.
The total free electron fraction is then written schematically as
\begin{equation}     
    x_e^{\rm tot}(z) = 
    x_e^{\rm rec}(z) + x_{e,{\rm H}}(z) + x_{e,{\rm He}}(z).
    \label{eq:xtot} 
\end{equation}
This construction is implemented in the thermodynamics module of the Boltzmann code \texttt{CLASS}~\cite{Lesgourgues:2011re,Diego_Blas_2011}.
The residual post-recombination contribution, $x_e^{\rm rec}(z)$, is computed using \texttt{HyRec-2}~\cite{PhysRevD.102.083517}; we supply the modified hydrogen contribution, $x_{e,{\rm H}}(z)$, while the helium contribution is incorporated internally by \texttt{CLASS}.

The resulting $x_e^{\rm tot}(z)$ determines the Thomson optical depth,
\begin{equation} 
    \tau_{\rm reio}(z_{\rm max}) = 
    \int_0^{z_{\rm max}} dz\, \frac{c\,\sigma_T\,n_e^{\rm tot}(z)} {(1+z)H(z)}, 
    \label{eq:tau_zmax} 
\end{equation} 
where $c$ is the speed of light, $\sigma_T$ is the Thomson cross section, and $n_e^{\rm tot}(z)$ is the physical free electron number density corresponding to $x_e^{\rm tot}(z)$.
Thus, $\tau_{\rm reio}$ is a derived quantity fixed by the total free electron history.

Our upper limit $z_{\rm max}$, following the \texttt{CLASS} convention, is identified with the redshift corresponding to the global minimum of $x_e^{\rm tot}(z)$ in the internally sampled history.
The reported $\tau_{\rm reio}$ is obtained by integrating the Thomson scattering rate from today up to this redshift.

From the same thermodynamic history, \texttt{CLASS} computes the Thomson scattering rate and visibility function used in the line-of-sight calculation of the CMB temperature and polarization anisotropies.
We adopt the high-accuracy configuration specified in Ref.~\cite{Calabrese_2025}.

\section{Statistical analysis and results}
\label{sec:analysis_results}

We use the model of Sec.~\ref{sec:ionization_model} to test how much additional optical depth DM decay can supply while remaining consistent with cosmological and reionization observations.
We compare a reference model with Gompertzian reionization and varying $\sum m_\nu$ to the extension that also includes DM decay, using the same joint likelihood and identical priors for the parameters common to both models.
After describing the parameters and data, we examine the representative case $m_\chi=1~\mathrm{GeV}$ to relate the posterior shifts to the CMB spectra and free electron history.
We then extend the analysis across a range of DM masses to constrain their conditional lifetime, and assess the implications for the CMB--BAO discrepancy.

\subsection{Parameters and priors}

We vary the cosmological parameters
$\boldsymbol{\theta}_{\rm cosmo}=\{n_s,~\ln(10^{10}A_s),~H_0,~\omega_\text{b},~\omega_\text{cdm},~\sum m_\nu\}$,
where $\omega_\text{b}$ and $\omega_\text{cdm}$ are the physical baryon and cold-dark-matter densities, $H_0$ is the Hubble constant, and $A_s$ and $n_s$ describe the amplitude and tilt of the primordial scalar power spectrum.
The sum of neutrino masses, $\sum m_\nu$, is free in both models.
The optical depth $\tau_{\rm reio}$ is derived from $x_e^{\rm tot}(z)$ through Eq.~\eqref{eq:tau_zmax} and is not sampled independently.

For reionization, we sample the transformed variables
$\boldsymbol{\theta}_{\rm reio}=\{\ln\alpha,~3/(2\beta)\}$,
which describe the timing and shape of the Gompertzian hydrogen baseline.

In the decay extension, we fix $m_\chi$ in each run and sample the decay rate, $\boldsymbol{\theta}_{\rm DM}=\{\Gamma_\chi\}$, together with the cosmological and reionization parameters.
We repeat this analysis over the mass grid in Table~\ref{tab:priors} and derive the lifetime $\tau_\chi=\Gamma_\chi^{-1}$ from each decay rate posterior to present constraints in the $(m_\chi,\tau_\chi)$ plane.

Table~\ref{tab:priors} lists the sampled, fixed, and derived parameters.
Unless otherwise stated, the priors are flat in the sampled variables over the listed ranges; in particular, the DM prior is flat in $\Gamma_\chi$.

\begin{table}[tbp]
\centering
\caption{Sampled, fixed, and derived parameters used in the analysis.
Flat priors are shown in square brackets.}
\renewcommand{\arraystretch}{1.12}
\setlength{\tabcolsep}{10pt}
\begin{tabular}{lll}
\toprule
\textbf{Sector} & \textbf{Parameter} & \textbf{Prior} \\
\midrule

\multirow{6}{*}{Cosmology}
& $n_s$ & $[0.92,1.00]$ \\
& $\ln(10^{10}A_s)$ & $[2.6,3.5]$ \\
& $H_0$ & $[61,73]~{\rm km\,s^{-1}\,Mpc^{-1}}$ \\
& $\omega_b$ & $[0.017,0.027]$ \\
& $\omega_{\rm cdm}$ & $[0.09,0.15]$ \\
& $\sum m_\nu$ & $[0,1.5]~{\rm eV}$ \\

\midrule

\multirow{3}{*}{Reionization}
& $\ln\alpha$ & $[-2.3,-1.7]$ \\
& $3/(2\beta)$ & $[0,1]$ \\
& $\tau_{\rm reio}$ & \textbf{derived} \\

\midrule

\multirow{3}{*}{DM decay}
& $m_\chi$
& \begin{tabular}[c]{@{}l@{}}
  \textbf{fixed} to $2,3,5,10,30,100,300~{\rm MeV}$\\
  and $1,3,7~{\rm GeV}$
  \end{tabular} \\
& $\Gamma_\chi$ & $[0,10^{-23}]~{\rm s}^{-1}$ \\
& $\tau_\chi$ & \textbf{derived} \\

\bottomrule
\end{tabular}
\label{tab:priors}
\end{table}

\subsection{Data, likelihoods, and posterior sampling}
\label{sec:data}
We constrain both models with a joint likelihood combining CMB anisotropies, CMB lensing, BAO measurements, and quasar damping-wing observations.
The CMB data constrain the free electron history and cosmological parameters, BAO measurements constrain the late-time expansion, and the quasar measurements directly constrain the neutral hydrogen fraction during the late stages of reionization.

For the primary CMB anisotropies, we use the Planck 2018 low-$\ell$ TT likelihood~\cite{2020A&A...641A...5P} together with the \texttt{SRoll2} low-$\ell$ EE likelihood~\cite{2019A&A...629A..38D}.
At high $\ell$, we follow the combination in Ref.~\cite{Louis_2025}: the ACT DR6 CMB-only TT, TE, and EE likelihood is combined with the Planck high-$\ell$ TT, TE, and EE likelihood~\cite{2021OJAp....4E...8E}, retaining the Planck spectra up to $\ell_{\rm max}=1000$ for TT and $\ell_{\rm max}=600$ for TE and EE.
ACT DR6 supplies the small-scale TT, TE, and EE information.

For CMB lensing, we use the combination of ACT DR6 and Planck PR4/NPIPE lensing measurements~\cite{Qu_2024,MacCrann_2024}.
ACT DR6 provides high-significance small-scale lensing information~\cite{Qu_2024,MacCrann_2024,Madhavacheril_2024,Farren_2024}, complemented by the larger-area Planck PR4/NPIPE reconstruction from reprocessed Planck maps~\cite{Carron_2022}.
The joint likelihood uses their complementary survey footprints and angular-scale sensitivities.

We complement the CMB likelihoods with BAO measurements from DESI DR2~\cite{tr6y-kpc6,2wwn-xjm5}.
These data help break the geometrical degeneracies of the CMB anisotropy and lensing measurements by constraining the late-time expansion history~\cite{Loverde:2024nfi}.

Following Refs.~\cite{Montero-Camacho:2024dzs,Montero-Camacho:2026ohy}, we include an astrophysical likelihood that compares the predicted neutral hydrogen fraction with high-redshift quasar Lyman-$\alpha$ damping-wing measurements~\cite{10.1093/mnras/stac825,10.1093/mnras/stae1080,Spina_2024,Durovcikova_2024}.
The relatively broad measurement uncertainties make these constraints conservative, while the well-constrained quasar systemic redshifts reduce one potential source of bias in the damping-wing analysis.
Figure~\ref{fig:ionization_history} and Table~\ref{tab:reionization_constraints} also include other external EoR measurements for comparison; only the quasar damping-wing measurements enter the astrophysical likelihood.

We embed the modified \texttt{CLASS} calculation in \texttt{Cobaya}~\cite{Torrado_2021} and explore the posterior with its Metropolis--Hastings Markov chain Monte Carlo (MCMC) sampler, running the decay extension independently at each fixed $m_\chi$.
Each evaluation uses the sampled cosmological and reionization parameters and decay rate to construct the free electron history and predict the observables entering the joint likelihood.
We use fast dragging between parameter blocks with different computational costs to improve sampling efficiency~\cite{PhysRevD.66.103511,PhysRevD.87.103529}.
We monitor convergence with the Gelman--Rubin statistic~\cite{10.1214/ss/1177011136} and require $R-1<0.01$ for the final chains.
We analyze the converged chains using \texttt{GetDist}~\cite{Lewis:2019xzd}.

We report marginalized posterior means and central $68\%$ credible intervals for the scalar constraints below.
The reionization histories are reconstructed and summarized by posterior medians and central $68\%$ and $95\%$ equal-tailed intervals, while the illustrative CMB spectra are evaluated at the posterior-mean parameter values.
Best-fit $\chi^2$ values are quoted separately to distinguish improvements in fit from shifts in the marginalized posteriors.

\subsection{Posterior constraints at a representative DM mass}
\label{sec:benchmark_results}

We choose $m_\chi=1~\mathrm{GeV}$ as a representative benchmark to illustrate the effects of DM decay on the inferred cosmological parameters and reionization history.
At this mass, DM decay raises the inferred optical depth but produces only a small improvement in the fit to the full data combination of Sec.~\ref{sec:data}.  
The best-fit value changes from $\chi_{\rm Gomp}^2=822.26$ in the reference model to $\chi^2_{\rm DM}(1\,{\rm GeV})=821.60$ in the decay extension, corresponding to $\Delta\chi^2=\chi^2_{\rm DM}(1\,{\rm GeV})-\chi^2_{\rm Gomp}=-0.66$.
For reference, $\chi^2_{\rm tanh} \approx 850$ for the same data, with a substantially larger contribution from the quasar damping-wing likelihood.
Given the additional decay parameter, this improvement is not statistically significant and provides no evidence for DM decay.
The main effect is instead a change in the allowed free electron histories and the associated parameter degeneracies.

\begin{figure}[tbp]
    \centering
    \includegraphics[width=0.6\textwidth]{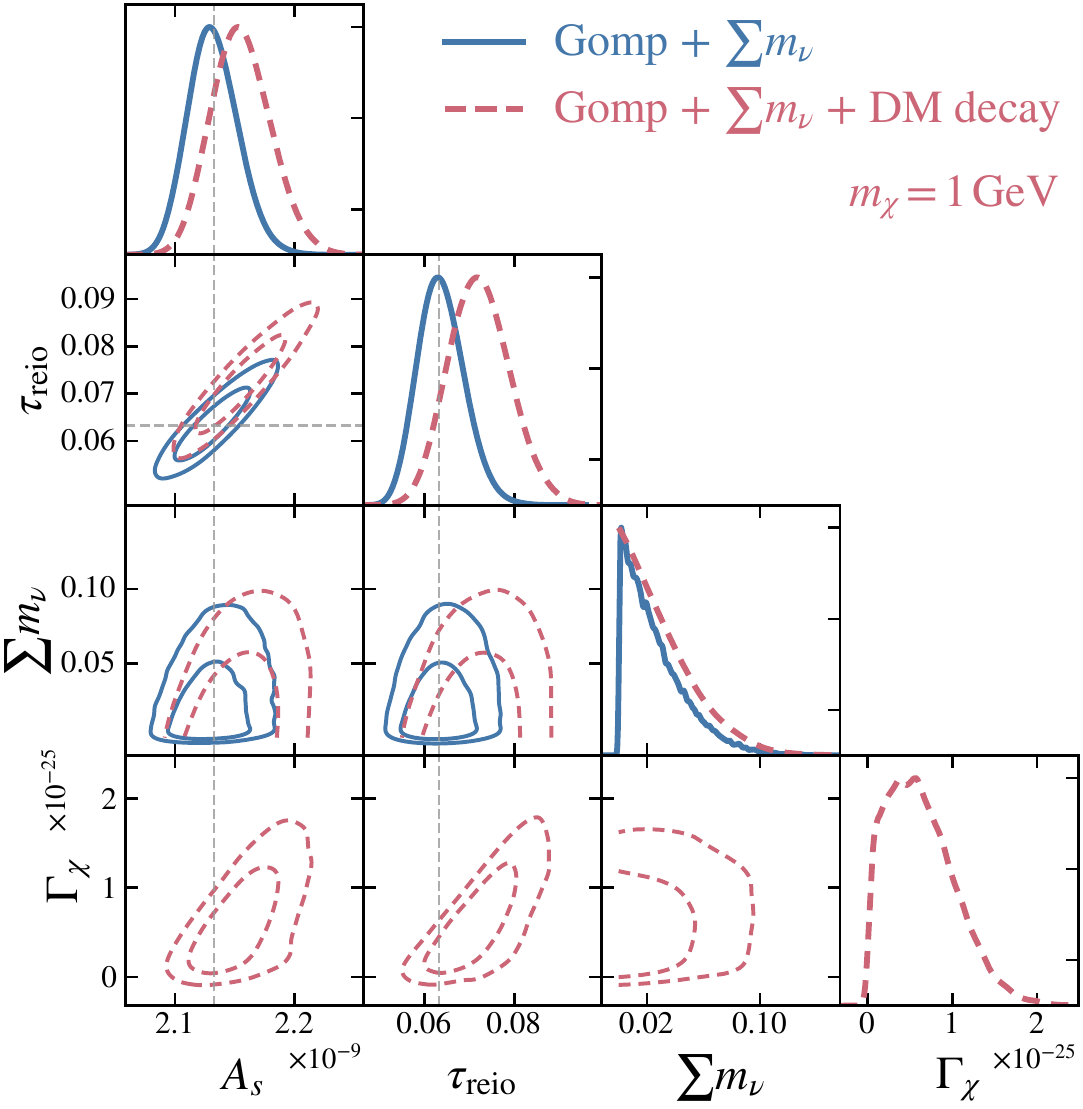}
    \caption{Marginalized posterior distributions for the reference model with Gompertzian reionization and varying $\sum m_\nu$, compared with its DM decay extension at fixed $m_\chi=1~\mathrm{GeV}$.
    The blue solid contours show the reference model, while the red dashed contours show the model including DM decay.
    The parameters shown are the primordial amplitude $A_s$, the reionization optical depth $\tau_{\rm reio}$, the sum of neutrino masses $\sum m_\nu$, and the DM decay rate $\Gamma_\chi$.
    The contours denote the $68\%$ and $95\%$ credible regions.
    The grey dashed lines indicate the corresponding $\Lambda$CDM parameter values taken from the P-ACT-LB column of Table 5 in Ref.~\cite{Louis_2025}.}
    \label{fig:triangle_plot}
\end{figure}

Figure~\ref{fig:triangle_plot} shows the marginalized constraints on $\tau_{\rm reio}$, $A_s$, $\sum m_\nu$, and $\Gamma_\chi$.
With a flat prior on $\Gamma_\chi$, nonzero decay rates allow larger $\tau_{\rm reio}$ and $A_s$, shifting the optical depth constraint from $\tau_{\rm reio}=0.0638^{+0.0046}_{-0.0057}$ in the reference model to $\tau_{\rm reio}=0.0722^{+0.0063}_{-0.0073}$ in the decay extension. The additional ionization from DM decay raises the optical depth, while a corresponding increase in $A_s$ approximately preserves the CMB amplitude combination $A_s e^{-2\tau_{\rm reio}}$.

The posterior for $\sum m_\nu$ remains concentrated near its lower prior boundary in both models, although the decay extension allows a broader tail toward larger masses.
We quantify this change through the posterior probability above the oscillation lower bound, $\sum m_\nu=0.06~{\rm eV}$~\cite{Esteban:2024eli}:
For the reference model, we find $P(\sum m_\nu \geq 0.06~{\rm eV}) = 8.5\%$, whereas for the $m_\chi=1~\mathrm{GeV}$ DM decay extension, this probability increases to $13.1\%$.
DM decay therefore shifts some posterior weight into the region compatible with oscillation measurements, but most of the posterior remains below $0.06~{\rm eV}$.

To assess the sensitivity of our results to the decay rate prior, we reweight the existing MCMC chains, originally sampled with a flat prior on $\Gamma_\chi$, to estimate the posterior under a log-uniform prior, $\pi_{\log}(\Gamma_\chi)\propto 1/\Gamma_\chi$, over $[\Gamma_{\min},\Gamma_{\max}]$.
We consider two representative lower cutoffs, $\Gamma_{\min}=10^{-27}\,\mathrm{s}^{-1}$ and $5\times10^{-27}\,\mathrm{s}^{-1}$, retaining the original upper cutoff for each DM mass and leaving all other priors unchanged.
Extending the prior to substantially lower decay rates can make the reweighting estimates unstable, as the $1/\Gamma_\chi$ weighting amplifies the contribution of samples with very small decay rates, which are sparsely represented in the existing chains.
Neither cutoff is assumed to be physically preferred.

For the $m_\chi=1\,\mathrm{GeV}$ model, the marginalized mean optical depth decreases from $0.0722$ under the flat prior to $0.0677$ and $0.0689$ for log-uniform priors with $\Gamma_{\min}=10^{-27}~\mathrm{s}^{-1}$ and $5\times10^{-27}~\mathrm{s}^{-1}$, respectively. 
The probability $P(\sum m_\nu\geq 0.06~\mathrm{eV})$ decreases from $13.1\%$ to $11.2\%$ and $11.6\%$, respectively, remaining above the $8.5\%$ reference model value. 
These results demonstrate the prior dependence of the marginalized constraints. 
The corresponding reweighted optical depth means across the full DM mass range are summarized in Table~\ref{tab:mass_constraints}.

\subsection{CMB signatures and the origin of the additional optical depth}
\label{sec:history_results}

\begin{figure}[tbp]
    \centering
    \includegraphics[width=\textwidth]{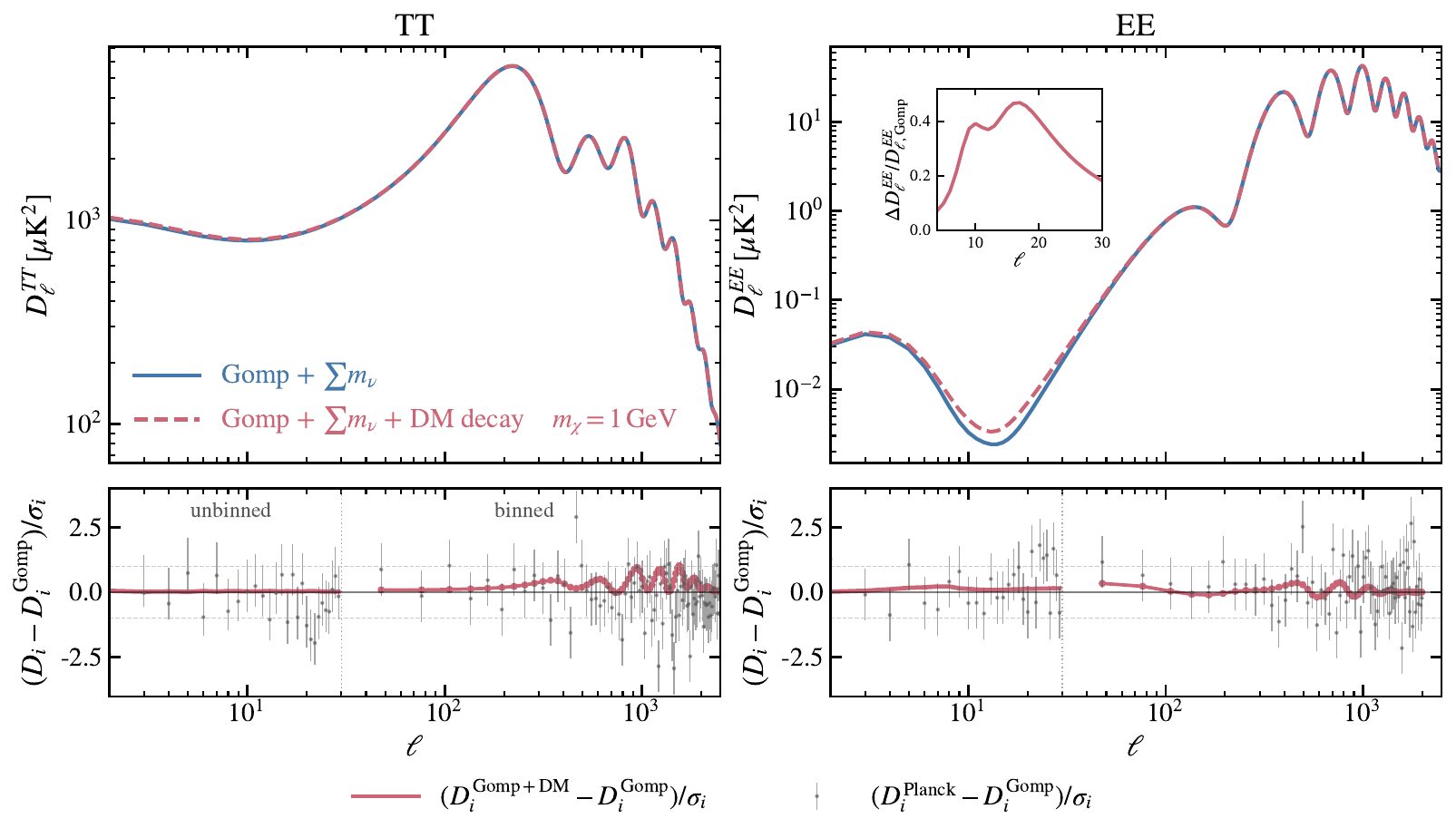}
    \caption{
    CMB temperature and $E$-mode polarization power spectra for the reference model with Gompertzian reionization and varying $\sum m_\nu$, and its DM decay extension at $m_\chi=1~\mathrm{GeV}$, evaluated at their respective posterior-mean parameter values.
    The upper panels show the TT and EE spectra, with $D_\ell^{XX}\equiv \ell(\ell+1)C_\ell^{XX}/(2\pi)$ for $X\in\{T,E\}$.
    The inset in the EE panel enlarges the fractional difference $\Delta D_\ell^{EE}/D_{\ell,\mathrm{Gomp}}^{EE}$ at low multipoles, where $\Delta D_\ell^{EE}\equiv D_{\ell,\mathrm{Gomp+DM}}^{EE}-D_{\ell,\mathrm{Gomp}}^{EE}$.
    In the lower panels, the red curves show the difference between the two model spectra normalized by the corresponding observational uncertainty, $(D_i^{\mathrm{Gomp+DM}}-D_i^{\mathrm{Gomp}})/\sigma_i$, while the grey points show the Planck data residuals relative to the reference model, $(D_i^{\mathrm{Planck}}-D_i^{\mathrm{Gomp}})/\sigma_i$.
    Here, $i$ labels an individual multipole for the unbinned low-$\ell$ data and a multipole bin for the binned higher-$\ell$ data. 
    For asymmetric Planck uncertainties, we take $\sigma_i=(\sigma_i^-+\sigma_i^+)/2$ for the normalization, while retaining the original asymmetric error bars in the plot.
    The vertical grey dotted line at $\ell=30$ marks the transition from the unbinned to the binned data.
    }
    \label{fig:cmb_tt_ee_spectra}
\end{figure}

Figure~\ref{fig:cmb_tt_ee_spectra} compares the TT and EE spectra at the respective posterior-mean parameter values of the two models.
The comparison shows how the higher inferred optical depth affects the CMB spectra.

The TT spectra remain nearly indistinguishable over the full multipole range, as do the high-$\ell$ EE spectra.
At high multipoles, the compensating shifts in $A_s$ and $\tau_{\rm reio}$ preserve approximately the same amplitude of the recombination-era temperature and polarization spectra.
The larger inferred optical depth produces a more visible enhancement of the low-$\ell$ EE spectrum by strengthening the large-scale reionization signal.
Even there, the lower panels show that the spectral differences are generally small compared with the corresponding Planck uncertainties.

\begin{figure}[tbp]
    \centering
    \includegraphics[width=\textwidth]{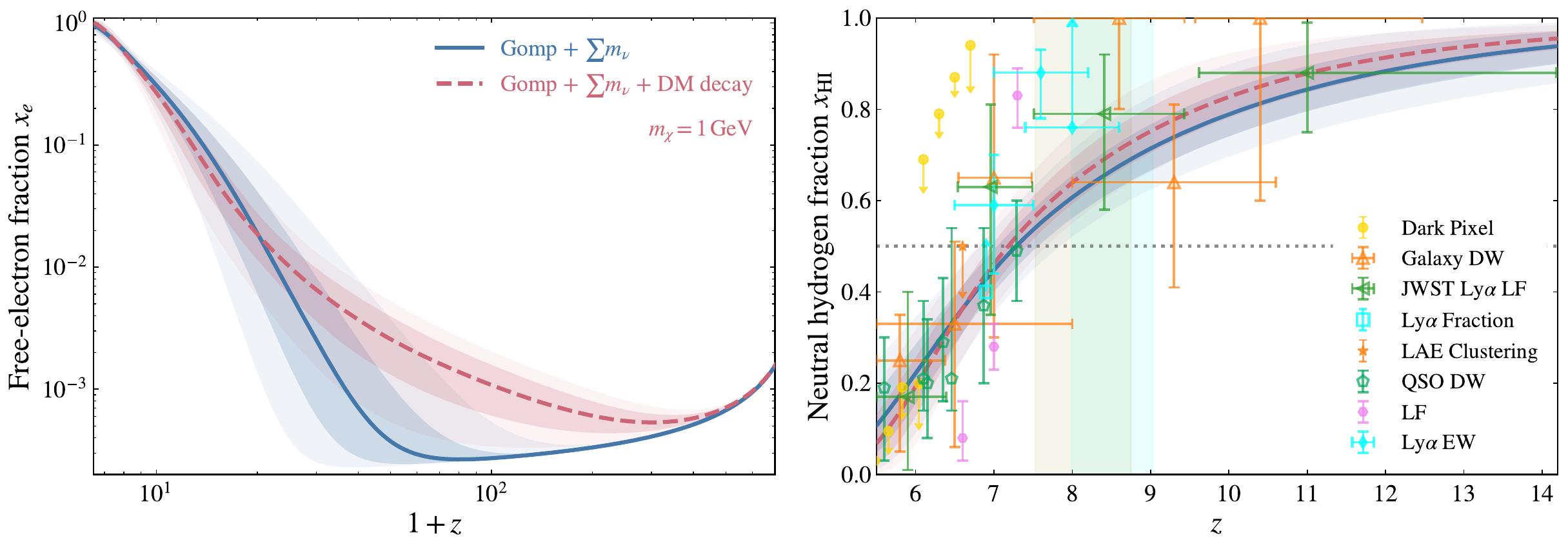}
    \caption{
    Reconstructed free electron and neutral hydrogen histories for the reference model with Gompertzian reionization and varying $\sum m_\nu$, and its DM decay extension at $m_\chi=1~\mathrm{GeV}$.
    The blue solid and red dashed curves represent the reference model and DM decay extension, respectively.
    The central curves denote the posterior medians, while the darker and lighter shaded regions show the central $68\%$ and $95\%$ equal-tailed credible intervals, corresponding to the $16$th--$84$th and $2.5$th--$97.5$th posterior percentiles, respectively.
    The left panel shows the free electron fraction $x_e$ as a function of $1+z$, illustrating the high-redshift ionization tail produced by DM decay.
    The right panel shows the corresponding neutral hydrogen fraction $x_{\rm HI}$ during the EoR, together with a compilation of astrophysical constraints.
    Only the quasar damping-wing measurements are included in the astrophysical likelihood used for parameter inference, whereas the remaining measurements are shown for comparison.
    The horizontal grey dotted line marks $x_{\rm HI}=0.5$.
    The vertical shaded bands show CMB optical depth constraints translated, assuming the conventional tanh prescription, into constraints on the midpoint redshift $z_{\rm re}$, defined by $x_{\rm HI}(z_{\rm re})=0.5$, and are displayed for reference only.
    The external constraints and their references are summarized in Table~\ref{tab:reionization_constraints}.
    }
    \label{fig:ionization_history}
\end{figure}

The reconstructed histories in Figure~\ref{fig:ionization_history} identify the redshifts responsible for the extra optical depth.
In the left panel, DM decay produces a broad ionization tail throughout the cosmic dark ages.
The absolute increase in $x_e$ is small, but it is substantial relative to the residual free electron fraction of the reference model and persists over a wide redshift interval.
This extended component makes an appreciable cumulative contribution to the optical depth integral in Eq.~\eqref{eq:tau_zmax}.

During the EoR, the posterior median of the decay extension has a slightly larger neutral hydrogen fraction over part of the transition, as shown in the right panel.
This corresponds to somewhat later astrophysical reionization, although the credible regions overlap substantially and the apparent delay is not statistically significant.
Together, the two panels indicate that the increase in $\tau_{\rm reio}$ arises primarily from the extended ionization tail in the dark ages.

The right panel also compares the histories with constraints from dark-pixel fractions, galaxy and quasar damping wings, Ly$\alpha$-emitter fractions and clustering, Ly$\alpha$ equivalent-width distributions, and Ly$\alpha$ luminosity functions.
As specified in Sec.~\ref{sec:data}, only the quasar damping-wing measurements enter the astrophysical likelihood.
The vertical bands show CMB optical depth constraints translated into the midpoint redshift $z_{\rm re}$, defined by $x_{\rm HI}(z_{\rm re})=0.5$, under the conventional tanh prescription; these bands are also for comparison only.
The measurements and their references are listed in Table~\ref{tab:reionization_constraints}.
Both models remain broadly consistent with the external constraints within the observational uncertainties over the redshift range shown.

\subsection{DM mass dependence and lifetime constraints}
\label{sec:mass_results}

The fixed-mass analyses give the marginalized decay rate and optical depth constraints listed in Table~\ref{tab:mass_constraints}, which also lists the reweighted mean optical depths under two log-uniform priors.
All masses yield only modest reductions in $\chi^2$ relative to the reference value $\chi^2_{\rm Gomp}=822.26$.
The lowest value is $\chi^2_{\rm DM}=821.55$ at $m_\chi=10\,{\rm MeV}$, almost identical to $821.56$ at $m_\chi=30\,{\rm MeV}$.
The best-fit $\chi^2$ values across the mass grid span less than $0.53$, providing no statistically meaningful preference for a particular mass.
Under the flat prior, the optical depth shifts upward at every mass: its marginalized posterior mean ranges from $0.0676$ to $0.0726$, compared with $0.0638$ in the reference model.
For both tested log-uniform priors, the reweighted means remain above the reference model value, but the increase is smaller than under the flat prior.

\begin{table}[tbp]
\centering
\small
\caption{Marginalized constraints on the reionization optical depth and DM decay rate for Gomp+$\sum m_\nu$ and Gomp+$\sum m_\nu$ + DM decay. 
The decay rate column and the flat prior optical depth column report posterior means and central $68\%$ credible intervals. 
The last two columns give the reweighted posterior means of $\tau_{\rm reio}$ under log-uniform priors with $\Gamma_{\min}=10^{-27}\,\mathrm{s}^{-1}$ and $5\times10^{-27}\,\mathrm{s}^{-1}$, respectively.
The upper cutoff $\Gamma_{\max}$ is retained from the original analysis for each DM mass. All constraints use the joint likelihood described in Sec.~\ref{sec:data}.}
\label{tab:mass_constraints}

\renewcommand{\arraystretch}{1.2}
\setlength{\tabcolsep}{5pt}

\begin{tabular}{ccccc}
\toprule

\multicolumn{5}{c}{
    \textbf{Gomp+$\sum m_\nu$}
}
\\[3pt]

\multicolumn{5}{c}{
    $\langle\tau_{\rm reio}\rangle
    =0.0638^{+0.0046}_{-0.0057}$
}
\\[4pt]

\midrule

\multicolumn{5}{c}{
    \textbf{Gomp+$\sum m_\nu$ + DM decay}
}
\\[6pt]

$m_\chi$
&
\shortstack[c]{
    $\Gamma_\chi/(10^{-25}\,\mathrm{s}^{-1})$\\
    Flat prior
}
&
\shortstack[c]{
    $\langle\tau_{\rm reio}\rangle$\\
    Flat prior
}
&
\shortstack[c]{
    $\langle\tau_{\rm reio}\rangle$\\
    Log-uniform prior\\
    $\Gamma_{\min}=10^{-27}\,\mathrm{s}^{-1}$
}
&
\shortstack[c]{
    $\langle\tau_{\rm reio}\rangle$\\
    Log-uniform prior\\
    $\Gamma_{\min}=5\times10^{-27}\,\mathrm{s}^{-1}$
}
\\[2pt]

\cmidrule{1-5}

$2\,\mathrm{MeV}$
& $1.15^{+0.37}_{-1.10}$
& $0.0700^{+0.0058}_{-0.0066}$
& $0.0662$
& $0.0669$
\\

$3\,\mathrm{MeV}$
& $0.95^{+0.29}_{-0.89}$
& $0.0676^{+0.0050}_{-0.0062}$
& $0.0650$
& $0.0655$
\\

$5\,\mathrm{MeV}$
& $0.60^{+0.19}_{-0.55}$
& $0.0691^{+0.0059}_{-0.0066}$
& $0.0657$
& $0.0666$
\\

$10\,\mathrm{MeV}$
& $0.39^{+0.14}_{-0.32}$
& $0.0719^{+0.0061}_{-0.0068}$
& $0.0680$
& $0.0696$
\\

$30\,\mathrm{MeV}$
& $0.22^{+0.08}_{-0.19}$
& $0.0725\pm0.0067$
& $0.0687$
& $0.0709$
\\

$100\,\mathrm{MeV}$
& $0.19^{+0.06}_{-0.17}$
& $0.0726^{+0.0063}_{-0.0074}$
& $0.0688$
& $0.0711$
\\

$300\,\mathrm{MeV}$
& $0.28^{+0.09}_{-0.25}$
& $0.0720^{+0.0062}_{-0.0072}$
& $0.0681$
& $0.0699$
\\

$1\,\mathrm{GeV}$
& $0.64^{+0.22}_{-0.56}$
& $0.0722^{+0.0063}_{-0.0073}$
& $0.0677$
& $0.0689$
\\

$3\,\mathrm{GeV}$
& $1.08^{+0.35}_{-0.98}$
& $0.0705^{+0.0058}_{-0.0067}$
& $0.0664$
& $0.0673$
\\

$7\,\mathrm{GeV}$
& $1.10^{+0.35}_{-1.00}$
& $0.0692^{+0.0056}_{-0.0067}$
& $0.0656$
& $0.0663$
\\

\bottomrule
\end{tabular}
\end{table}

At fixed $m_\chi$, increasing $\Gamma_\chi$ strengthens the extended ionization tail, producing the positive correlation with $\tau_{\rm reio}$ seen in Figure~\ref{fig:triangle_plot}.
The mass dependence has a different origin.
At fixed $\Gamma_\chi$, the injected power in Eq.~\eqref{eq:dE_inj} has no explicit dependence on $m_\chi$: the number density scales as $n_\chi\propto m_\chi^{-1}$, while the energy released per decay scales as $E_{\rm decay}\propto m_\chi$.
The mass dependence therefore enters through the deposition efficiencies $f_\alpha(z,m_\chi)$ in Eq.~\eqref{eq:dE_dep}.
For injected electrons and positrons, Ref.~\cite{PhysRevD.95.023010} found an enhancement in the hydrogen ionization efficiency $f_{\rm ion}(z)$ for initial kinetic energies around $1$--$100\,{\rm MeV}$, because electrons in this range upscatter CMB photons to energies from tens of eV to the keV range, where they can efficiently ionize hydrogen.
For the two-body decay $\chi\rightarrow e^+e^-$, this enhanced electron-energy range corresponds broadly to DM masses from a few to a few hundred MeV.

The optical depths in Table~\ref{tab:mass_constraints} include both astrophysical reionization and DM decay.
Since $\Gamma_\chi$ and the reionization parameters vary jointly, a stronger ionization response at a given mass can be offset by a smaller inferred decay rate and adjustments to the astrophysical history.
The posterior means of $\tau_{\rm reio}$ therefore need not follow the mass dependence of the inferred decay rate.

\begin{figure}[!htb]
    \centering
    \includegraphics[width=0.7\textwidth]{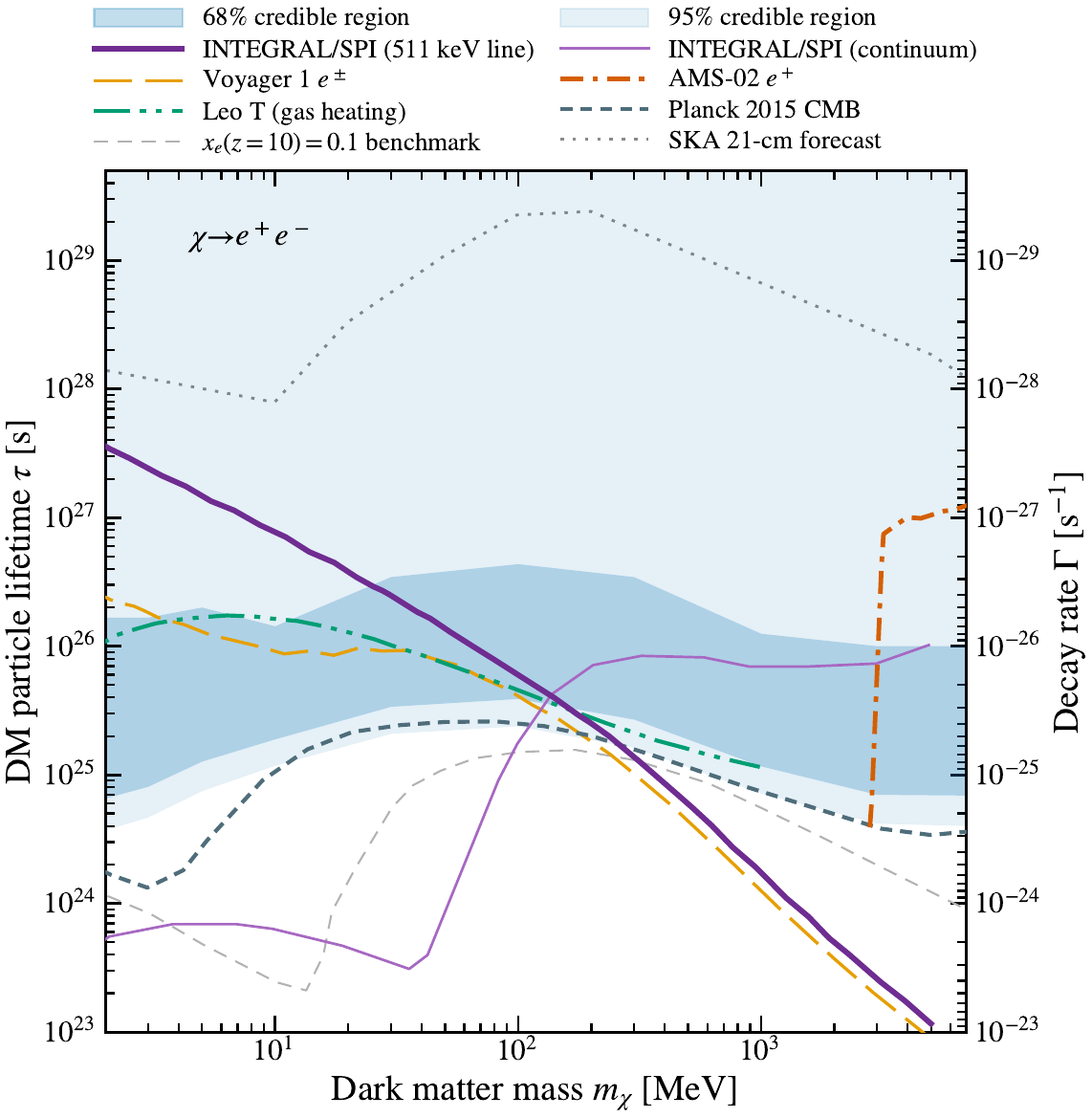}
    \caption{Marginalized DM lifetime constraints for $\chi\rightarrow e^{+}e^{-}$ as a function of $m_\chi$.
    The dark and light blue bands indicate the $68\%$ and $95\%$ credible regions obtained from the posterior for decay rate at each fixed DM mass.
    The right-hand axis shows the corresponding decay rate, $\Gamma_\chi=\tau_\chi^{-1}$.
    For comparison, we show the INTEGRAL/SPI constraints from the Galactic longitude profile of the 511 keV positron annihilation line~\cite{DeLaTorreLuque2024SPI511,DeLaTorreLuque2025SPI511Erratum} and the Galactic latitude profiles of diffuse X-ray and soft gamma-ray emission~\cite{Cirelli2023Xray,Cirelli2025XrayErratum}, 
    together with bounds from Voyager 1 $e^\pm$~\cite{DelaTorreLuque_2024}, 
    AMS-02 $e^+$~\cite{PhysRevLett.113.121102,PhysRevLett.119.021103}, 
    gas heating in Leo~T~\cite{PhysRevD.106.075007}, 
    and Planck 2015 CMB observations~\cite{PhysRevD.95.023010}. 
    The $x_e(z=10)=0.1$ curve marks the lifetime at which DM decay produces this free electron fraction, assuming instantaneous reionization at $z=10$~\cite{PhysRevD.94.063507}, and serves as a benchmark rather than an observational bound. 
    The grey dotted curve shows the projected sensitivity of future SKA 21-cm observations~\cite{Zhao_2026}.}
    \label{fig:dm_decay_lifetime_posterior}
\end{figure}

Figure~\ref{fig:dm_decay_lifetime_posterior} expresses the decay rate constraints as lifetimes, $\tau_\chi=\Gamma_\chi^{-1}$, for comparison with existing astrophysical and CMB bounds and future projections.

The lifetime constraints exhibit a non-monotonic dependence on the DM mass.
The lower boundary of the $95\%$ credible region moves toward longer lifetimes as $m_\chi$ increases from a few MeV, reaches a maximum at masses of order tens to hundreds of MeV, and then shifts back toward shorter lifetimes in the GeV range.
This trend is qualitatively consistent with the mass dependence of energy deposition described above; the detailed shape also reflects degeneracies with the cosmological and reionization parameters.

INTEGRAL/SPI observations provide stringent constraints from the Galactic 511 keV positron annihilation line~\cite{DeLaTorreLuque2024SPI511,DeLaTorreLuque2025SPI511Erratum} and diffuse X-ray and soft gamma-ray emission~\cite{Cirelli2023Xray,Cirelli2025XrayErratum}.
The 511 keV line constraint is particularly strong at low DM masses, where lower-energy positrons propagate over shorter distances before thermalization, resulting in more localized annihilation emission. 
At higher masses, the continuum constraint becomes more competitive as inverse Compton emission contributes more strongly within the INTEGRAL/SPI energy range.
These constraints depend on the assumed Galactic DM distribution and the propagation and radiative processes of the decay products.

Voyager 1 probes low-energy cosmic-ray electrons and positrons~\cite{DelaTorreLuque_2024}, although its lifetime constraints are weaker than those from the SPI 511 keV line over the mass range considered here. 
AMS-02 positron measurements provide complementary constraints at higher DM masses~\cite{PhysRevLett.113.121102,PhysRevLett.119.021103}. 
Their sensitivity decreases toward lower masses as the injected positron energies fall below the observed energy range, although Galactic reacceleration can extend the sensitivity to lower DM masses~\cite{PhysRevLett.119.021103}.

The Leo~T constraint requires that heating from DM decay not exceed the gas cooling rate and is particularly strong in the MeV mass range~\cite{PhysRevD.106.075007}.
Compared with the Planck 2015 CMB constraint~\cite{PhysRevD.95.023010}, the lower boundary of our $95\%$ credible region lies at longer lifetimes at low masses and is broadly comparable at higher masses.

The $x_e(z=10)=0.1$ curve marks the lifetime at which DM decay produces this free electron fraction under the assumption of instantaneous reionization at $z=10$~\cite{PhysRevD.94.063507}.
It is a benchmark for substantial pre-reionization ionization, not an observational exclusion limit.
The SKA projection indicates the additional sensitivity that future 21-cm observations could provide~\cite{Zhao_2026}.

\subsection{Implications for the CMB--BAO discrepancy}
\label{sec:optical_depth_implications}

The mass scan shows that DM decay consistently raises the inferred optical depth, but the size of the shift remains limited.
Ref.~\cite{6r54-8lv4} found that $\tau_{\rm reio}\simeq0.09$ could remove the cosmological preference for neutrino masses below the oscillation lower bound, while $\tau_{\rm reio}\simeq0.08$ could reduce all of the discrepancies considered in that analysis to below $2\sigma$.
For every mass studied here, both the flat-prior posterior mean and the upper boundary of its central $68\%$ credible interval remain below $\tau_{\rm reio}=0.08$; the reweighted mean optical depths under both tested log-uniform priors also remain below this benchmark.
Relative to those benchmarks, the posterior-favored increase is insufficient to substantially alleviate the discrepancy.
This is consistent with the $1~\mathrm{GeV}$ result: the posterior probability above the oscillation lower bound increases, but most of the neutrino-mass posterior still lies below it.

\section{Conclusions}

In this work, we have investigated whether electromagnetic energy injection from decaying DM can provide the additional Thomson optical depth suggested by the optical-depth interpretation of the emerging discrepancy between CMB and BAO measurements~\cite{6vd2-rbfn}.
Rather than varying $\tau_{\rm reio}$ as an independent phenomenological parameter, we derived it self-consistently from the evolution of the free electron fraction. 
We adopted a Gompertzian reionization model as the astrophysical hydrogen baseline~\cite{Montero-Camacho:2024dzs,Montero-Camacho:2026ohy} and supplemented it with energy injection from DM decaying through the benchmark channel $\chi\rightarrow e^{+}e^{-}$.
The DM-induced ionization and heating were calculated with \texttt{DarkHistory}, from which we extracted the resulting modification to the hydrogen-sector free electron fraction. 
This modification was combined with the astrophysical hydrogen baseline and supplied to \texttt{CLASS}, where the helium contribution was incorporated to construct the total free electron fraction and compute the CMB observables.
We then constrained the model using a joint likelihood combining the latest ACT+Planck CMB temperature and polarization anisotropies, CMB lensing, BAO measurements, and quasar damping-wing constraints on the late stages of reionization.
This framework therefore tests whether a higher optical depth can arise from a concrete ionization mechanism while remaining consistent with both cosmological and astrophysical observations.  

For the representative case $m_\chi=1\,{\rm GeV}$, the inclusion of DM decay produces only a small improvement in the best-fit $\chi^2$, from $\chi^2_{\rm Gomp}=822.26$ to $\chi^2_{\rm DM}(1\,\rm GeV)=821.60$, corresponding to $\Delta\chi^2=-0.66$.
Across the fixed-mass scan, the best-fit $\chi^2$ values span less than $0.53$, with nearly identical minima of $821.55$ and $821.56$ at $m_\chi=10$ and $30\,{\rm MeV}$, respectively.
Given the additional decay parameter, these modest improvements are not statistically significant and provide no evidence for DM decay or for a preferred DM mass.
Nevertheless, nonzero decay rates open a degeneracy direction toward larger $\tau_{\rm reio}$ and $A_s$. 
For $m_\chi=1\,{\rm GeV}$, under a flat prior on $\Gamma_\chi$, the marginalized optical depth constraint shifts from $\tau_{\rm reio}=0.0638^{+0.0046}_{-0.0057}$ in the reference model to $\tau_{\rm reio}=0.0722^{+0.0063}_{-0.0073}$ when DM decay is included. 
The corresponding increase in $A_s$ approximately preserves the CMB amplitude combination $A_s e^{-2\tau_{\rm reio}}$, while shifting additional posterior weight toward larger neutrino masses.
In particular, the probability of satisfying the oscillation lower bound~\cite{Esteban:2024eli}, $P(\sum m_\nu\geq0.06\,{\rm eV})$, increases from 8.5\% to 13.1\%. 
Reweighting the existing chains under the two tested log-uniform priors lowers the mean optical depth to $0.0677$ and $0.0689$, and the corresponding probability to $11.2\%$ and $11.6\%$, respectively.
Thus, the DM decay extension moves the neutrino-mass inference in the direction favored by oscillation measurements, although most of the posterior remains below the physical lower bound and the shift does not constitute a resolution of the discrepancy.  

The reconstructed histories clarify the physical origin of the optical depth enhancement.
DM decay generates a broad ionization tail extending through the cosmic dark ages and toward the onset of astrophysical reionization (see Figure \ref{fig:ionization_history}).
Although the absolute increase in the free electron fraction at any individual redshift is modest, it persists over a wide redshift interval and therefore makes an appreciable cumulative contribution to the line-of-sight optical depth.
The enhancement in $\tau_{\rm reio}$ is consequently driven primarily by this extended high-redshift component rather than by a shift in the timing of astrophysical reionization.
Indeed, the posterior median of the DM decay extension exhibits a slightly larger neutral fraction over part of the reionization epoch, corresponding to a somewhat later astrophysical transition, although the credible regions of the reference model and the DM decay extension overlap substantially and this difference is not statistically significant. 
Both reionization histories remain broadly consistent with the external reionization measurements shown for comparison.  

Extending the analysis over the full mass range $m_\chi=2\,{\rm MeV}$--$7\,{\rm GeV}$, we find that the marginalized posterior means under the flat prior range from $\tau_{\rm reio}=0.0676$ to $0.0726$, compared with $\tau_{\rm reio}=0.0638$ in the reference model.
Reweighting to either of the two tested log-uniform priors lowers the mean optical depth at every mass, although the reweighted means remain above the reference model value.
Previous analyses have shown that an optical depth of $\tau_{\rm reio}\simeq0.09$ could make the cosmological preference for neutrino masses below the oscillation lower bound statistically insignificant~\cite{6vd2-rbfn,6r54-8lv4}, while values as low as $\tau_{\rm reio}\simeq0.08$ could reduce the related cosmological discrepancies, including the reported exclusion of $\Lambda$CDM at more than $2\sigma$, to below $2\sigma$~\cite{6r54-8lv4}.
For all masses studied here, however, both the posterior mean and the upper boundary of the central 68\% credible interval remain below 0.08. 
Within the decay channel, mass range, Gompertzian reionization model, and data combination considered in this work, DM decay therefore shifts the cosmological inference in the favorable direction but does not generate enough additional optical depth to substantially alleviate the CMB–BAO discrepancy.

Independently of its impact on the optical-depth interpretation of the CMB--BAO discrepancy, this analysis yields constraints on the lifetime of DM decaying into electron–positron pairs. 
The lifetime constraints exhibit a non-monotonic dependence on the DM mass, becoming strongest at masses of order tens to hundreds of MeV and weakening toward both lower masses and the GeV regime. 
This behavior reflects the energy- and redshift-dependent propagation, cooling, and deposition of the injected electrons and positrons.
The resulting cosmological lifetime constraints can be compared with those from INTEGRAL/SPI observations of the 511 keV positron annihilation line~\cite{DeLaTorreLuque2024SPI511,DeLaTorreLuque2025SPI511Erratum} and diffuse continuum emission~\cite{Cirelli2023Xray,Cirelli2025XrayErratum}, Voyager 1~\cite{DelaTorreLuque_2024} and AMS-02~\cite{PhysRevLett.113.121102,PhysRevLett.119.021103} cosmic-ray measurements, and gas heating in the Leo~T dwarf galaxy~\cite{PhysRevD.106.075007}. 
The INTEGRAL/SPI limits are particularly stringent at lower DM masses, while AMS-02 provides the strongest constraints at the high-mass end. 
Our cosmological constraints are broadly comparable to or stronger than previous CMB limits over parts of the mass range~\cite{PhysRevD.95.023010}.
Future 21-cm observations, in particular with the SKA, could provide substantial additional sensitivity to the ionization and heating histories generated by DM decay~\cite{Zhao_2026}.

Overall, our results therefore show that decaying DM can produce an upward shift in $\tau_{\rm reio}$ while remaining consistent with the data considered here, but that the magnitude of the allowed shift is insufficient to fully address the tension indicated by current CMB--BAO comparisons.

\section*{Acknowledgments} We are grateful to Xin-Chen Duan for valuable discussions. The numerical computations in this work were supported by the National Natural Science Foundation of China (No. 12588101). They were also supported by the National Key Research and Development Program of China (No. 2022YFF0503304). This work is supported by NSFC (grant No. 12603005), the National Key Research and Development Program of China under grant number 2023YFA1605600, the Basic and Frontier Research Project of PCL (grant No. 2025QYB012), and the Major Key Project of PCL.

\section*{Data and Code Availability}
The code and MCMC chains used in this work are available from Yu-Ning Wang upon reasonable request.

\newpage

\appendix

\section{External Reionization Constraints}

\begingroup

\setlength{\LTpre}{0pt}
\setlength{\LTpost}{0pt}
\setlength{\LTleft}{0pt}
\setlength{\LTright}{0pt}

\setlength{\LTcapwidth}{\textwidth}

\setlength{\tabcolsep}{2.4pt}

\renewcommand{\arraystretch}{0.8}
\setlength{\extrarowheight}{0.5pt}

\newlength{\papersepsize}
\setlength{\papersepsize}{9pt}

\newcommand{\papersep}{%
    \addlinespace[\papersepsize]%
}

\newcommand{\labelrule}{%
    \specialrule{0.25pt}{5pt}{5pt}%
}

\renewcommand{\tabularxcolumn}[1]{m{#1}}

\newcommand{\headerstrut}{%
    \rule[-1.25ex]{0pt}{4.4ex}%
}

\newcommand{\onepapergap}[1]{%
    \raisebox{\dimexpr-\papersepsize/2\relax}[0pt][0pt]{#1}%
}

\newcommand{\twopapergaps}[1]{%
    \raisebox{\dimexpr-\papersepsize\relax}[0pt][0pt]{#1}%
}

\newcommand{\threepapergaps}[1]{%
    \raisebox{\dimexpr-3\papersepsize/2\relax}[0pt][0pt]{#1}%
}

% ============================================================
% Table
% ============================================================

\begin{xltabular}{\textwidth}{
    @{}
    >{\scriptsize\raggedright\arraybackslash}m{0.18\textwidth}
    @{\hspace{10pt}}
    >{\scriptsize\centering\arraybackslash}m{0.16\textwidth}
    @{\hspace{3pt}}
    >{\scriptsize\centering\arraybackslash}X
    @{\hspace{3pt}}
    >{\scriptsize\centering\arraybackslash}m{0.28\textwidth}
    @{\hspace{3pt}}
    >{\scriptsize\centering\arraybackslash}m{0.08\textwidth}
    @{}
}

\caption{
External constraints on reionization shown in Fig.~\ref{fig:ionization_history}.
The CMB measurements constrain the midpoint redshift $z_{\rm re}$, corresponding to $\langle x_{\rm HI}\rangle=0.5$ under the standard tanh reionization parametrization. 
The other measurements constrain the volume-averaged neutral hydrogen fraction. 
The statistical convention associated with each constraint follows that adopted in the corresponding reference.
The quasar damping-wing measurements, marked with an asterisk, are included in the reionization likelihood; all other measurements are shown for comparison.
}
\label{tab:reionization_constraints}\\

\toprule

\headerstrut
\textbf{Probe}
&
\textbf{$z$}
&
\textbf{Constraint}
&
\textbf{Statistical summary}
&
\textbf{Ref.}
\\

\midrule

\endfirsthead

\toprule

\headerstrut
\textbf{Probe}
&
\textbf{$z$}
&
\textbf{Constraint}
&
\textbf{Statistical summary}
&
\textbf{Ref.}
\\

\midrule

\endhead

\endfoot

\bottomrule
\endlastfoot

% ============================================================
% CMB midpoint-redshift constraints
% ============================================================

\multirow[c]{2}{=}{%
    \onepapergap{CMB optical depth}
}
&
$z_{\rm re}=8.14\pm0.61$
&
$\langle x_{\rm HI}\rangle=0.5$
&
$1\sigma$ confidence interval
&
\cite{Pagano_2020}
\\

\papersep

&
$z_{\rm re}=8.51\pm0.52$
&
$\langle x_{\rm HI}\rangle=0.5$
&
$1\sigma$ confidence interval
&
\cite{deBelsunce:2021mec}
\\

\labelrule

% ============================================================
% Dark Pixel
% ============================================================

\multirow[c]{8}{=}{%
    \raggedright
    \onepapergap{Dark Pixel}
}
&
$\langle z\rangle = 5.481$
&
$\langle x_{\rm HI}\rangle \leq 0.030 + 0.048$
&
\multirow[c]{4}{=}{%
    \centering
    Fiducial $1\sigma$ upper limits
}
&
\multirow[c]{4}{*}{%
    \cite{10.1093/mnras/staf1862}
}
\\

&
$\langle z\rangle = 5.654$
&
$\langle x_{\rm HI}\rangle \leq 0.095 + 0.037$
& &
\\

&
$\langle z\rangle = 5.831$
&
$\langle x_{\rm HI}\rangle \leq 0.191 + 0.056$
& &
\\

&
$\langle z\rangle = 6.043$
&
$\langle x_{\rm HI}\rangle \leq 0.199 + 0.087$
& &
\\

\papersep

&
$z = 6.1$
&
$\langle x_{\rm HI}\rangle < 0.69\pm0.06$
&
\multirow[c]{4}{=}{%
    \centering
    Upper limits with $1\sigma$
    confidence intervals
}
&
\multirow[c]{4}{*}{%
    \cite{Jin_2023}
}
\\

&
$z = 6.3$
&
$\langle x_{\rm HI}\rangle < 0.79\pm0.04$
& &
\\

&
$z = 6.5$
&
$\langle x_{\rm HI}\rangle < 0.87\pm0.03$
& &
\\

&
$z = 6.7$
&
$\langle x_{\rm HI}\rangle
< 0.94^{+0.06}_{-0.09}$
& &
\\

\labelrule

% ============================================================
% Galaxy DW
% ============================================================

\multirow[c]{6}{=}{%
    \raggedright
    \onepapergap{Galaxy DW}
}
&
$\langle z\rangle=5.8$
&
$\langle x_{\rm HI}\rangle
=0.25^{+0.10}_{-0.20}$
&
\multirow[c]{4}{=}{%
    \centering
    Posterior modes with $1\sigma$
    highest-posterior-density intervals
}
&
\multirow[c]{4}{*}{%
    \cite{Umeda_2026}
}
\\

&
$\langle z\rangle=7.0$
&
$\langle x_{\rm HI}\rangle
=0.65^{+0.27}_{-0.35}$
& &
\\

&
$\langle z\rangle=8.6$
&
$\langle x_{\rm HI}\rangle
=1.00^{+0.00}_{-0.20}$
& &
\\

&
$\langle z\rangle=10.4$
&
$\langle x_{\rm HI}\rangle
=1.00^{+0.00}_{-0.40}$
& &
\\

\papersep

&
$\langle z\rangle=6.5$
&
$\langle x_{\rm HI}\rangle
=0.33^{+0.18}_{-0.27}$
&
\multirow[c]{2}{=}{%
    \centering
    $1\sigma$ credible intervals
}
&
\multirow[c]{2}{*}{%
    \cite{2026A&A...705A.114M}
}
\\

&
$\langle z\rangle=9.3$
&
$\langle x_{\rm HI}\rangle
=0.64^{+0.17}_{-0.23}$
& &
\\

\labelrule

% ============================================================
% JWST Ly-alpha LF
% ============================================================

\multirow[c]{4}{=}{%
    \raggedright
    JWST Ly$\alpha$ LF
}
&
$\langle z\rangle=5.90$
&
$\langle x_{\rm HI}\rangle
=0.17^{+0.23}_{-0.16}$
&
\multirow[c]{4}{=}{%
    \centering
    Posterior modes with $1\sigma$
    highest-posterior-density intervals
}
&
\multirow[c]{4}{*}{%
    \cite{Kageura_2025}
}
\\

&
$\langle z\rangle=6.96$
&
$\langle x_{\rm HI}\rangle
=0.63^{+0.18}_{-0.28}$
& &
\\

&
$\langle z\rangle=8.41$
&
$\langle x_{\rm HI}\rangle
=0.79^{+0.13}_{-0.21}$
& &
\\

&
$\langle z\rangle=11.00$
&
$\langle x_{\rm HI}\rangle
=0.88^{+0.11}_{-0.13}$
& &
\\

\labelrule

% ============================================================
% Ly-alpha Fraction
% ============================================================

Ly$\alpha$ fraction
&
$z\simeq7$
&
$\langle x_{\rm HI}\rangle \gtrsim 0.4$
&
$1\sigma$ confidence-level lower limit
&
\cite{10.1093/mnras/stu2089}
\\

\labelrule

% ============================================================
% LAE Clustering
% ============================================================

LAE clustering
&
$z\simeq6.6$
&
$\langle x_{\rm HI}\rangle \lesssim 0.5$
&
$1\sigma$ upper limit
&
\cite{10.1093/mnras/stv1751}
\\

\labelrule

% ============================================================
% QSO damping wings
% ============================================================

\multirow[c]{7}{=}{%
    \raggedright
    \threepapergaps{QSO DW$^{\ast}$}
}
&
$z=7.29$
&
$\langle x_{\rm HI}\rangle
=0.49^{+0.11}_{-0.11}$
&
$1\sigma$ credible interval
&
\cite{10.1093/mnras/stac825}
\\

\papersep

&
$z=6.15$
&
$\langle x_{\rm HI}\rangle
=0.20^{+0.14}_{-0.12}$
&
\multirow[c]{2}{=}{%
    \centering
    Posterior medians with $1\sigma$
    credible intervals
}
&
\multirow[c]{2}{*}{%
    \cite{10.1093/mnras/stae1080}
}
\\

&
$z=6.35$
&
$\langle x_{\rm HI}\rangle
=0.29^{+0.14}_{-0.13}$
& &
\\

\papersep

&
$z=5.6$
&
$\langle x_{\rm HI}\rangle
=0.19^{+0.11}_{-0.16}$
&
$2\sigma$ uncertainty
&
\cite{Spina_2024}
\\

\papersep

&
$\langle z\rangle=6.10$
&
$\langle x_{\rm HI}\rangle
=0.21^{+0.17}_{-0.07}$
&
\multirow[c]{3}{=}{%
    \centering
    Maximum-likelihood values with
    16th--84th percentiles of the
    marginalized posteriors
}
&
\multirow[c]{3}{*}{%
    \cite{Durovcikova_2024}
}
\\

&
$\langle z\rangle=6.46$
&
$\langle x_{\rm HI}\rangle
=0.21^{+0.33}_{-0.07}$
& &
\\

&
$\langle z\rangle=6.87$
&
$\langle x_{\rm HI}\rangle
=0.37^{+0.17}_{-0.17}$
& &
\\

\labelrule

% ============================================================
% Ly-alpha Luminosity Function
% ============================================================

\multirow[c]{3}{=}{%
    \raggedright
    Ly$\alpha$ LF
}
&
$z=6.6$
&
$\langle x_{\rm HI}\rangle
=0.08^{+0.08}_{-0.05}$
&
\multirow[c]{3}{=}{%
    \centering
    Posterior medians with $1\sigma$
    credible intervals
}
&
\multirow[c]{3}{*}{%
    \cite{Morales_2021}
}
\\

&
$z=7.0$
&
$\langle x_{\rm HI}\rangle
=0.28\pm0.05$
& &
\\

&
$z=7.3$
&
$\langle x_{\rm HI}\rangle
=0.83^{+0.06}_{-0.07}$
& &
\\

\labelrule

% ============================================================
% Ly-alpha Equivalent Width
% ============================================================

\multirow[c]{3}{=}{%
    \raggedright
    \twopapergaps{Ly$\alpha$ EW}
}
&
$z\simeq7$
&
$\langle x_{\rm HI}\rangle
=0.59^{+0.11}_{-0.15}$
&
Posterior median with
16th--84th percentile bounds
&
\cite{2018ApJ...856....2M}
\\

\papersep

&
$z\simeq7.6$
&
$\langle x_{\rm HI}\rangle
=0.88^{+0.05}_{-0.10}$
&
$1\sigma$ confidence interval
&
\cite{2019ApJ...878...12H}
\\

\papersep

&
$z\simeq8.0\pm0.6$
&
$\langle x_{\rm HI}\rangle >0.76$
&
One-sided $1\sigma$
posterior lower limit
&
\cite{10.1093/mnras/stz632}
\\

\end{xltabular}

\endgroup

\clearpage
\bibliographystyle{JHEP} 
\bibliography{reference} 

\end{document}